\documentclass[useAS,usenatbib]{mnras}
\usepackage{natbib}
\usepackage{graphicx}
\usepackage{graphicx,graphics,amsmath,aas_macros}
\usepackage{multirow}
\usepackage{subfig}
\usepackage{amssymb}
\usepackage{hyperref}
\usepackage{textcomp}

\def\ergcms{{\rm ~erg~cm^{-2}~s^{-1}}}

\newcommand{\suzaku}{Suzaku}
\newcommand{\astro}{AstroSat}
\newcommand{\rxte}{{RXTE}}
\newcommand{\bep}{ BeppoSAX}
\newcommand{\gin}{ Ginga}

\newcommand{\src}{4U 1538--522}

\def\Msun{\hbox{$\rm ~M_{\odot}$}}

\begin{document}
\title[]{Probing the Cyclotron line characteristics of 4U 1538--522 using \astro-LAXPC }
\author[xx]{Varun$^{1}$\thanks{Contact e-mail: varun@rri.res.in},  Chandreyee Maitra$^{2}$, Pragati Pradhan$^{3,4}$, Harsha Raichur$^{5}$ and Biswajit Paul$^{1}$\\
1. Department of Astronomy and Astrophysics, Raman Research Institute, Banglore-560080, India \\
2. Max Planck Institute For Extraterrestrial Physics, 85748 Garching, Germany \\
3. Department of Astronomy and Astrophysics, Pennnsylvania State University, Pennnsylvania, 16802, US\\
4. St Joseph’s College, Singmari, Darjeeling, 734104, India\\
5. Nordita, KTH Royal Institute of Technology and Stockholm University, Rosalagstullsbacken, 23, SE-10691 Stockholm}

\date{Accepted.....; Received .....}

\maketitle

\begin{abstract}

We report the first report on cyclotron line studies with the LAXPC instrument onboard \astro\ of the High mass X-ray Binary pulsar 4U 1538--52. During the observation of source which spanned about one day with a net exposure of 50 ks, the source X-ray flux remained constant. Pulse profile is double peaked in low energy range and has a single peak in high energy range, the transition taking place around the cyclotron line energy of the source. Cyclotron Scattering Feature (CRSF) is detected  at $\sim$22 keV with a very high significance in phase averaged spectrum. It is one of the highest signal to noise ratio detection of CRSF for this source. We performed detailed pulse phase resolved spectral analysis with 10 independent phase bins. We report the results of pulse phase resolved spectroscopy of the continuum and CRSF parameters. The cyclotron line parameters show pulse phase dependence over the entire phase with a CRSF energy variation of $\sim$ 13$\%$ which is in agreement with previous studies. We also confirm the increase in the centroid energy of the CRSF observed between the 1996--2004 ({\it \rxte}) and the 2012 ({\it \suzaku}) observations, reinforcing that the increase was a long-term change. 
\end{abstract}

\begin{keywords}
binaries: general-- stars: Neutron--star: pulsars-- radiation mechanism: non-thermal-- individual: 4U 1538--522; 
\end{keywords}

\section{Introduction}
Neutron Star High-mass X-ray binaries (HMXBs) have strong magnetic fields of the order of $10^{11}-10^{13}$ G. The emitted X-ray radiation is broadband in nature and is generated by multiple and interdependent processes near the neutron star surface. The strong magnetic fields in HMXBs play a crucial role in the generated X-ray radiation. Due to the non-spherical nature of the emission region, and the scattering cross sections which are altered by the magnetic fields, different emission beam patterns are produced: the fan-beam type emission at high accretion rates, and the pencil-beam type emission at low accretion rates. The distinction between the high and low accretion rate regimes is believed to be determined by the critical luminosity, which also depends on the magnetic field of the neutron star \citep{basko-sunyaev1976,becker2012,2015MNRAS.447.1847M}. Another interesting feature is the presence of Cyclotron Resonance Scattering Feature (CRSF)
usually seen in absorption against the continuum spectrum in many of these sources. CRSFs are formed by cyclotron resonant scattering of the X-ray photons in the presence of the strong magnetic field of the neutron star.
The centroid energy of the CRSF is related to the magnetic field given by the ``12-B-12'' rule given by \begin{equation}
	E_{\mathrm{cyc}} = \frac{11.57\,\mathrm{keV}}{1+z} \times B_{12}
	\label{eqn:12b12}
\end{equation}
where $B_{12}$ is the magnetic field in units of $10^{12}$\,G and z
is the gravitational redshift in the scattering region for neutron star. An estimated mass of $1.0\pm0.10\Msun$ \citep{2011ApJ...730...25R} and radius of 10 km gives $z\sim0.15$ for neutron star in this system. Pulse phase dependence of the CRSF parameters provide important information on the line forming region at different viewing angles,
like the plasma temperature and the optical depth of the line forming region. Most importantly, it provides crucial clues on the geometry of the line forming region: the emission beam pattern, and the magnetic field configuration of the neutron star \citep[][for a recent review on CRSFs]{2017JApA...38...50M}.

4U 1538--522 is a wind-fed persistent X-ray binary formed by a massive (17\Msun) B0Iab supergiant \citep{1992MNRAS.256..631R} and a neutron star. At the time of its discovery, the neutron star had a spin period of $\sim528$ sec \citep{1977ApJ...216L..11B} and it was in spin down phase for more than a decade. Since then source underwent two torque reversals, in 1998 \citep{1997ApJ...488..413R} and 2009 \citep{2013ApJ...777...61H}. The spin period is currently $\sim 527$ s. The orbit of the binary system has a high inclination ($67^{\circ}$) \citep{2015A&A...577A.130F}, and has an orbital period period of 3.75 days with an X-ray eclipse lasting $\sim$ 0.6 days \citep{1977ApJ...216L..11B,2006JApA...27..411M}. However, shape of orbit is still unknown for this system. Some authors have interpreted the orbit to be circular  \citep{1987ApJ...314..619M,1995A&A...303..497V} whereas others have found an eccentricity of 0.17-0.18 \citep{2000ApJ...542L.131C,2006JApA...27..411M}. \cite{2011ApJ...730...25R} have used both circular and elliptical orbit parameters for mass determination of the neutron star from optical light curves and radial velocity measurements. The system consists of supergiant donor that under-fills its Roche Lobe and is similar to other wind fed systems like Vela X-1 and 4U 1907+09 exhibiting short time variability in terms of dips and flaring activity.  The persistent X-ray luminosity is estimated to be about 2 $\times$ 10$^{36}$ erg/s for a distance of $\sim$ 6.4 kpc \citep{1977ApJ...216L..11B}.\\
The X-ray spectrum is  typical of an accreting HMXB pulsar with a power-law feature and an exponential turnover. A CRSF at $\sim$ 20 keV was discovered by \cite{1990ApJ...353..274C} using a {\it Ginga} observation. An additional absorption feature was identified in the spectrum at $\sim$ 51 keV using {\it BeppoSAX} observations \citep{2001ApJ...562..950R}, and was later confirmed as the first harmonic of the CRSF by \cite{2009A&A...508..395R}. \cite{2014ApJ...792...14H} reported the results of phase resolved analysis of the CRSF fundamental in 4U 1538--522 keV for the first time using a {\it \suzaku\ } observation. Noticeable variations in the CRSF parameters were observed. However, the pattern of variation could not be probed in detail as the parameters had large error bars associated with them, due to statistical limitations of the data.  \cite{2016MNRAS.458.2745H} reported that CRSF energy of \src\ has increased by $\sim1.5$ keV between 1996 and 2012. Further measurement of cyclotron line energy are required to confirm that this increase is secular. 

Here we report the first timing and spectral analysis results of 4U 1538--522 from a 50 ks \astro\, observation using the LAXPC detectors. Sect. 2 presents the observation and analysis, Sect. 3 the results, Sect. 4 the discussions and Sect. 5 the conclusions and summary. \\

\section{Observation \& Analysis} 

The Large Area X-ray Proportional Counter (LAXPC) onboard \astro\ consists of three identical Proportional Counters viz., LXP10, LXP20 and LXP30, with seven anodes arranged into five layers, and each filled with Xe gas at 2 atmosphere pressure. The geometric area of each proportional counter is $\sim$ $100 \times 36$ cm$^{2}$ \citep{2017ApJS..231...10A} with slightly different effective area for three different instruments. The broadband nature of the instrument with high time resolution capability makes it ideal to perform a detailed timing and spectral studies of HMXB pulsars and especially study CRSFs \citep{2013IJMPD..2241009P}. 4U 1538--522 was observed with \astro\, \citep{2006AdSpR..38.2989A,2014SPIE.9144E..1SS} on 12th and 13th July 2017 for $\sim$ 50 ks.  We report here the results of analysis using LAXPC data of the source taken in the event analysis (EA) mode. Data files were reduced with LAXPC data analysis and calibration software\footnote{\url{http://www.rri.res.in/~rripoc/POC.html}} version 1. Level1 products were processed to reduce level2 products from which we can produce light curves and spectral files. Further handling of light curves and spectra was done using tools in HEASOFT software suite, version 6.19. The time intervals when the source is occulted by earth and the satellite is passing through the South Atlantic Anomaly (SAA) region have been removed for the creation of light curves and spectral files. Average of the count rates during earth occultation, and the spectra acquired during earth occultation with the satellite being outside the SAA are used as background count rate and spectra in subsequent sections. \~.

\begin{figure}
\centering
\includegraphics[width=0.26\textwidth,angle=-90]{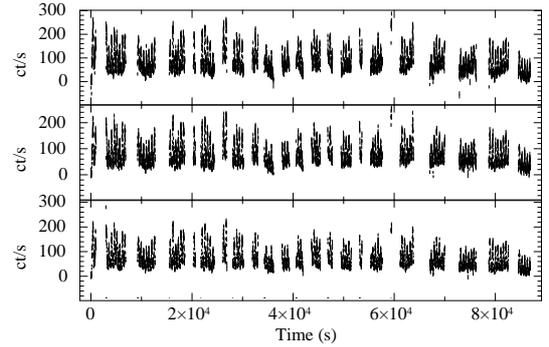}
\caption{Light curves of 4U 1538-52 in energy band 3.0-80.0 keV from 3 LAXPC detectors -- From top to bottom LXP10, LXP20, LXP30. Data gaps corresponds to times when satellite is passing through SAA, or source is occulted by earth. }
\label{lc123}
\end{figure}

\begin{figure}
\centering
\includegraphics[width=0.25\textwidth,angle=-90]{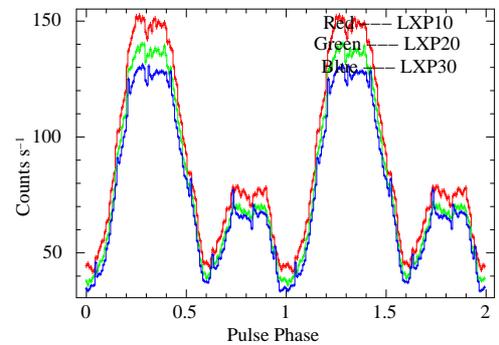}
\caption{X-ray Pulse profiles of 4U 1538-52 created from light curves folded with $P_{spin} = 527.10$ s. Pulse profiles from three detectors -- LXP10 (red), LXP20 (green), LXP30 (blue) are overlapped.}
\label{pp123}
\end{figure}

\begin{figure}
\centering
\includegraphics[width=9cm,height=8cm,angle=-90]{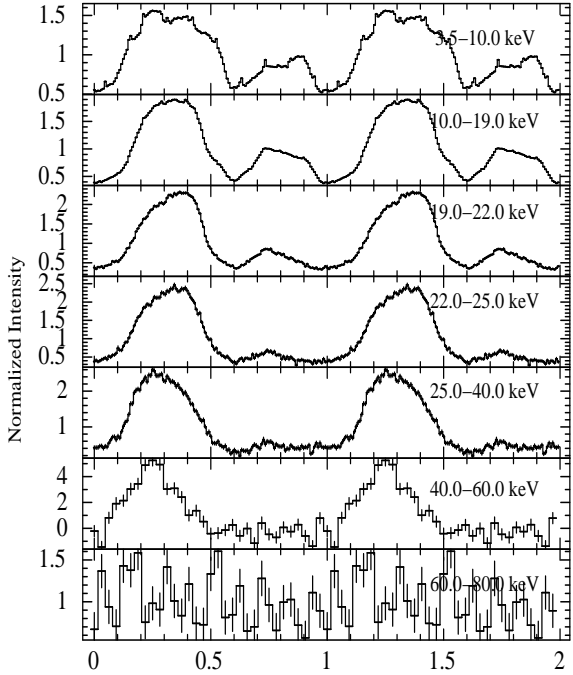}
\caption{Energy resolved pulse profiles from light curves of data added from 3 detectors. The 3.0-80.0 keV energy band has been divided into 7 finer energy bands.}
\label{enrpp2}
\end{figure}

\subsection{Timing analysis}
For the timing analysis, light curves from the entire observation were extracted with a bin size of 10 ms. Barycenter correction was applied to light curve using tool \texttt{as1bary} \footnote{\url{http://astrosat-ssc.iucaa.in/?q=data\_and\_analysis}} along with online tool AstroSat orbit file generator\footnote{\url{http://astrosat-ssc.iucaa.in:8080/orbitgen/}}. We have not done orbital correction on the photon arrival time as there is ambiguity about the orbital parameters of this source. Fig.~\ref{lc123} shows the light curves binned at 10 s for LXP10, LXP20 and LXP30 with average source count rates of 88.4,79.9  and 75.6 c/s respectively. Pulsations are clearly seen in the light curves of all the three detectors. We used the pulse folding and $\chi^{2}$-maximization method to determine the pulse period of the pulsar using the FTOOL \texttt{efsearch}. Pulsations were detected at  $527.06\pm0.11$ s using all the three detectors. Fig.~\ref{pp123} shows the background subtracted pulse profile from all the detectors integrated in the energy range of 3.0--80.0 keV. Pulse profiles show a double peaked structure with a primary and secondary peak. To investigate the energy dependence of the same, we extracted pulse profiles in seven energy bands from 3 to 80 keV, selecting two energy bands (19.0-22.0 and 22.0-25.0 keV) around cyclotron  absorption feature as shown in Figure \ref{enrpp2}. The pulse profiles are double peaked and exhibit significant evolution with energy.  The secondary peak (phase 0.6--1.0) decreases in strength until it disappears for energy $>$ 19 keV. The primary peak also becomes narrower with energy with pulse fraction increasing from 48\% to $\sim$ 98\% (for 40.0-60.0 keV). Pulse fraction in band for the low energy side of CRSF line center (19.0-22.0 keV) is higher (74\%) than to higher energy side (22.0-25.0 keV) where its value is 70\%.  Pulsations are detected up to $\sim$ 60 keV. In 60-80 keV energy band pulsations are not detected due to poor statistics in data.  The pattern of evolution of the pulse profiles with energy are consistent with the results from \gin\ \citep{1990ApJ...353..274C} ,\bep\ \citep{2001ApJ...562..950R}, and {\it \suzaku\ } \citep{2014ApJ...792...14H}.

\begin{figure}
\centering
\includegraphics[width=0.3\textwidth,angle=-90]{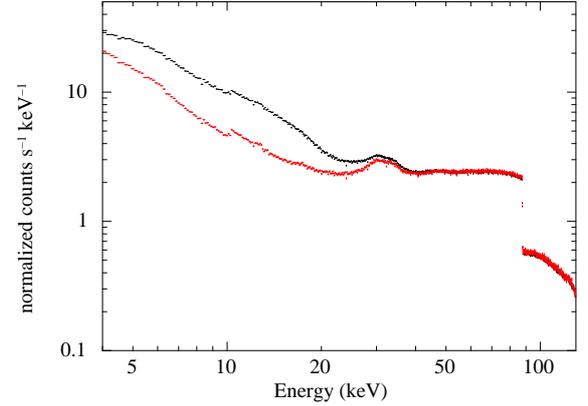}
\caption{Source and background spectra used for spectroscopy used in this work. }
\label{srcbkg}
\end{figure}

\begin{figure}
\centering
\includegraphics[width=0.3\textwidth,angle=-90]{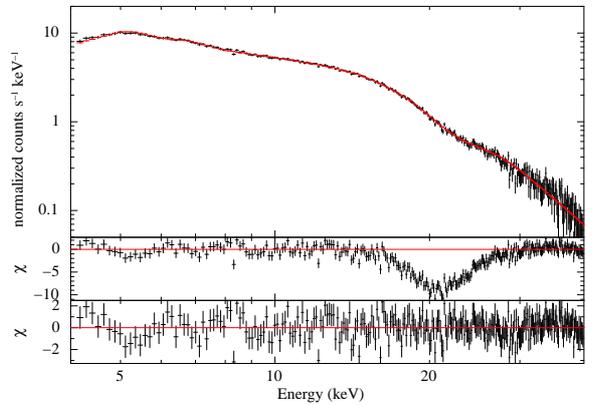}
\caption{Pulse-phase-averaged spectrum of 4U 1538-52 along with best fit model using LXP10 data. Top panel shows data (Black) and model (Red). Middle panel shows residuals when model does not contain component for CRSF. Bottom panel show residues when CRSF component is included in the model.}
\label{spec1}
\end{figure}

\subsection{Spectral analysis: Broadband spectroscopy}

 Spectra were extracted from all the 1024 channels which encompassed the full energy range of 3.0-80.0 keV. Total exposure of the source spectrum is $\sim50$ ks. The earth occulted spectrum was used as background, which is shown along with the spectrum during the source observation in Figure~\ref{srcbkg}. X-ray photons with energy higher than the K-shell energy of Xenon, i.e 34.0 keV, produce fluorescence emission which in some cases lead to simultaneous double events. Spectra of the fluorescent events among the double events are used for gain calibration. During this observation, due to higher gain in LXP20, some of the double events are not recorded within the preset window for fluorescent events. Therefore, LXP20 data was not used for spectroscopy. Data from LXP30 was also not used for spectroscopy as it has a continuously varying gain and decreasing effective area due to gas leakage \citep{2017ApJS..231...10A}.\\

The spectra of accretion powered pulsars are widely modeled phenomenologically by a power-law with quasi exponential high energy cutoffs of
 various functional forms ({\it XSPEC} models `highecut', `bknpow', `fdcut' and `newhcut'). Other models include a combination of two power-laws with different photon
indices but a common cutoff energy value called the Negative and Positive powerlaws with Exponential model ({\it XSPEC} model 
'NPEX' ), and a thermal Comptonization model ({\it XSPEC} model 'CompTT).  A careful modeling of the broadband continuum spectrum is crucial 
for obtaining consistent and physically reliable parameters for the CRSF \citep[see for example][]{muller2013}. We fit the spectra of 4U 1538--522 with all the above mentioned models. We found that two models `fdcut' and `newhcut' (a smoothed high energy cutoff model) provided better fits to the continuum spectrum compared to other models in terms of absence of residuals and the reduced $\chi^{2}$ values. Fdcut model is a powerlaw with a form of the exponential cutoff energy that resembles with the Fermi Dirac function. The mathematical details of model can be found from \cite{1986bookT}. Newhcut model is a modification of the high energy cutoff model, smoothed around the cutoff energy. Functional form of this model was given in \cite{2000ApJ...530..429B}. The constants in this model are calculated internally assuming the continuity of the intensity function and its derivative in the range of $E_C \pm \Delta E$.  This model has been used \citep{2015MNRAS.448..620J,2017MNRAS.470..713M,2018MNRAS.480L.136M} extensively for broad band spectra of many accreting pulsars with cyclotron line and $\Delta E$ is usually kept constant at 5.0 keV. To be consistent with the literature, we have also fixed it to 5.0 keV.\\
 
  A uniform systematic error across the entire energy band gave large contributions to the $\chi^{2}$ in the 4.0-10.0 keV range without any structures in the residuals in that energy range. To reduce the $\chi^{2}$, we have therefore added  a systematic error of 1\% in energy band 4.0-10.0 keV and 0.5\% in 10.0-40.0 keV to the spectra using the tool GRPPHA. The fit parameters were found to be largely the same if same systematics was used across the full band, albeit with a poorer spectral fit. A narrow and deep absorption feature at $\sim$ 22 keV was observed in the residuals, coincident with the CRSF known in this source. The feature was modeled with a Gaussian absorption feature ({\it XSPEC} model `Gabs'). A Fe K$\alpha$ line was also detected at 6.4 keV and was modeled by a Gaussian emission line. For the final spectral model, a power-law modified by `newhcut' and a `Gabs' absorption line for the CRSF was adopted. This is because the CRSF is better constrained in the `newhcut' model compared to the `fdcut'. Using `newhcut` as continuum and 'Gabs` as CRSF we obtain a fit with $\chi^{2}$ of 277.59 for 276 dof. When a fit is done without `gabs` component, we get a relatively poor fit with $\chi^{2}$ of 937.53 for 279 dof. This makes it one of highest signal to noise ratio detection of CRSF for this source. In order to strengthen the reliability of CRSF parameters we performed a contour analysis in {\it XSPEC} and found that CRSF energy is not correlated with $\Delta E$ parameter of the Newhcut model. Figure~\ref{spec1} shows the phase averaged spectrum with the best-fit model and the residuals with and without adding the CRSF component. The best-fit spectral parameters are reported in Table \ref{tab}.\\
\begin{table}
\caption{Best fitting phase averaged spectral parameters of 4U 1538--522. Errors quoted are for 90\% confidence range.}
\begin{tabular}{c c c}
\hline \hline
parameters & NEWHCUT & FDCUT\\
\hline
$\Gamma$  			& 0.96 $\pm$ 0.02 	& 0.87 $\pm$ 0.03 	\\
E-folding energy (keV) 		& 10.3$\pm$ 0.3  	& 5.0$\pm$ 0.5		\\
E-cut energy (keV) 		& 14.0$\pm$ 0.3 	& 24.0$\pm$ 1.5 	\\
E$_{CRSF}$ (keV) 		& 21.9$\pm$ 0.2   	& 22.7$\pm$ 0.3 	\\
$\sigma_{CRSF}$ (keV) 		& 2.31$\pm$ 0.3  	& 3.5$\pm$ 0.3		\\
$\tau_{CRSF}$ 			& 2.9$\pm$ 0.4		& 8$\pm$1		\\
Fe K$\alpha$ (keV) 		& 6.45$\pm$ 0.09 	& 6.40$\pm$0.09		\\
Fe K$\alpha$ EQW (keV) 		& 0.21$\pm$ 0.03 	& 0.21$\pm$0.03		\\
Flux  $^a$ (4.0-40.0 keV) 	& 1.102$\pm$ 0.006	& 1.10$\pm$ 0.01  	\\
Reduced $\chi^{2}$/d.o.f  	& 1.00/276		& 1.16/276		\\
With CRSF 			& 			& 			\\
Reduced $\chi^{2}$/d.of  	& 3.36/279		&6.29/279 		\\
Without CRSF 			& 			& 			\\
\hline
\end{tabular}\\
$^{a}$ - Observed flux in units of $10^{-9}$ $\ergcms$ in the mentioned energy band. \\
\label{tab}
\end{table}
\subsection{Pulse phase resolved spectroscopy}
\begin{figure}
\centering
\includegraphics[width=8cm,height=11cm]{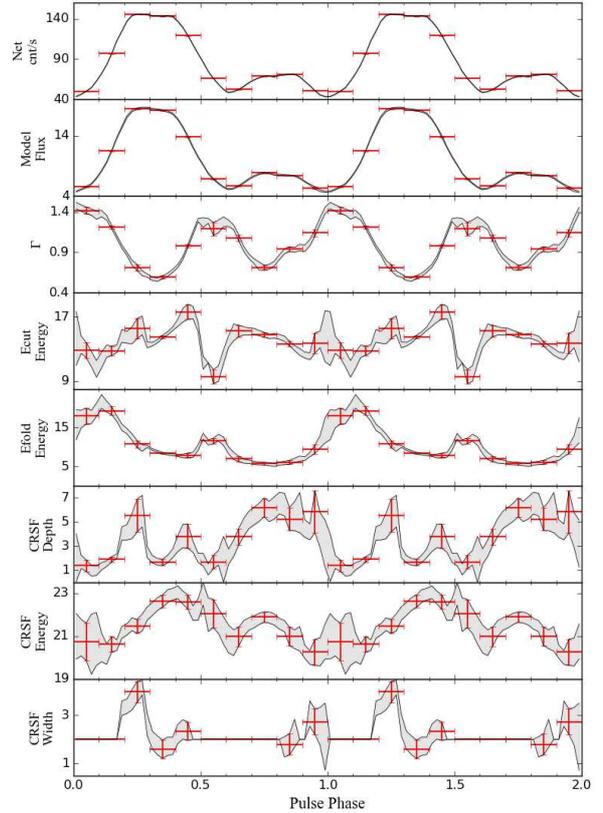}
\caption{ Phase resolved variations of the continuum and CRSF parameters.
Results from the independent phase bins are in black and of the sliding phase bins are in red.}
\label{pava}
\end{figure}

\begin{figure}
\centering
\includegraphics[width=0.3\textwidth,angle=-90]{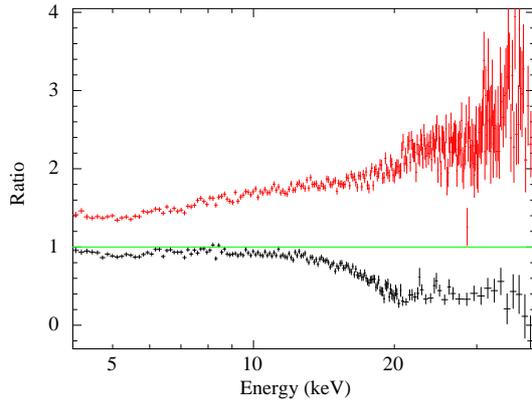}
\caption{ Ratio of the spectrum during pulse phase 0.3-0.4 and the phase averaged spectrum is shown in red, while the same for phase 0.7-0.8 is shown in black. The spectrum is softer during phase 0.7-0.8 with a deep CRSF.}
\label{cdep}
\end{figure}

In order to probe the changes in the continuum and the CRSF parameters with the rotational phase of the pulsar, spectra were extracted from 10 independent pulse phase bins.  
The same background spectra and response file as in case of phase averaged  spectral analysis were used.  We also extracted spectra with overlapping bins of phase width 0.1 each sliding with a phase interval of 0.02. We refer to this as sliding phase resolved spectroscopy. However, all the 50 spectra in sliding phase resolved spectroscopy are not statistically independent. Effectively it is a set of five different phase resolved analyses, each with ten independent bins. The sliding phase resolved spectroscopy confirms that the pulse phase dependence of the cyclotron line parameters is insensitive to the choice of phase zero. The upper and lower limits of the parameters obtained from the sliding phase resolved spectroscopy are shown in Figure \ref{pava} with shaded regions. Figure~\ref{pava} shows results of the phase resolved spectral analysis of the continuum and CRSF parameters. The spectrum is harder at the primary (phase 0.2-0.4 ) and secondary pulse peaks (phase 0.7-0.8) with lower values of $\Gamma$ and corresponding increase in $E_{cut}$ and $E_{fold}$ energies. The secondary peak is softer in general than the primary peak and is consistent with the energy dependence of the pulse profiles seen in Fig.~\ref{enrpp2}. The CRSF parameters show significant energy dependence with the phase as also observed in earlier works \citep{1990ApJ...353..274C,2014ApJ...792...14H}. The variation of all the CRSF parameters are however probed in detail for the first time, especially the CRSF depth. In Figure~\ref{cdep} we highlight the strong pulse phase dependence of the X-ray spectrum. Ratio of two of the pulse phase resolved spectra are shown here. The spectrum at phase 0.3-0.4 (shown in red) is harder than the phase averaged spectrum whereas at phase 0.7-0.8 (shown in black) it is softer with a strong CRSF feature around 22 keV. The CRSF energy ($E_{\mathrm cyc}$) is correlated with the pulse profile and rises to $\sim$ 23 keV near the peak of the primary peak (phase 0.4-0.5). This is consistent with that seen in earlier works \citep{1990ApJ...353..274C,2014ApJ...792...14H}. The overall variation in the CRSF energy measured from independent bins is 13 $\%$(20.00-23.02 keV). The CRSF is deeper near the secondary peak (phase 0.65-0.95) than the primary peak (phase 0.2-0.3), and lot of variation is detected near the primary pulse. It was not possible to constrain the CRSF width at the rise (phase 0.0-0.2) of the primary peak and during secondary peak (phase 0.5-0.8) and was frozen to the phase averaged value $\sim$ 2.0 keV instead for these phases. The overall pattern implies a complex pulse phase dependence of the CRSF parameters which are evident especially from the sliding phase resolved spectroscopy. The parameter variations are consistent with that obtained from the independent phase bins.

\section{Discussion and Conclusions}
We presented here the first results of \src\ using a LAXPC observation onboard AstroSat. We probed the timing and spectral properties of the source especially at hard X-rays and presented the most detailed results of pulse phase resolved spectral analysis available for the source till date.  

\cite{2016MNRAS.458.2745H} reported a long term increase in the CRSF centroid energy by $\sim$ 1.5\, keV between the 1996--2004 {\it RXTE} and 2012 {\it Suzaku} measurements. The authors obtained a linear increasing trend of 0.058$\pm$0.014 keV yr$^{-1}$. This would imply an increase of 0.29 keV between the  {\it \suzaku\ } and \astro\  measurements which would be impossible to detect given the error bars of the measurements. Our reported centroid energies of $E_{\mathrm cyc}=21.9\pm0.2$\,keV  with the `newhcut' continuum model and $E_{\mathrm cyc}=22.76\pm0.26$\,keV with the `fdcut' continuum model are however consistent with that obtained from the {\it \suzaku\ } data while all the earlier {\it RXTE} measurements were below 21\, keV. Our results are comparable to results obtained by \cite{2016MNRAS.458.2745H}. The obtained result from our work further strengthens the claim of a long term increase in  $E_{\mathrm cyc}$, rather than the {\it \suzaku\ }, measurement representing a short-term or a local increase in $E_{\mathrm cyc}$. The long term increase indicates a local change or reconfiguration in the line forming region, thereby resulting in an increased effective magnetic field strength \citep{2012MNRAS.420..720M}. However future observations separated by sufficiently large time gaps will be able to ascertain whether the trend is linearly increasing or the increase was due to a sudden change in the line-forming region which caused this change in $E_{cyc}$. \\

Our results of phase variation dependence of CRSF energy are in agreement with earlier works. \cite{1990ApJ...353..274C} studied CRSF energy variation with pulse phase using {\it Ginga} observation. They used power-law as a continuum component along with a gabs component for CRSF. They found a $14\%$ variation in CRSF energy with pulse phase. \cite{2014ApJ...792...14H} used {\it Suzaku} observations and found $10\%$ variation of same using highecut, fdcut and NPEX models. The variation using \astro\ LAXPC observation is $\sim 13 \%$.

The measured flux of \src\ in the energy range of 4.0--40.0 keV indicates a luminosity of $5.1\times10^{36}$ erg s$^{-1}$ at a distance of 6.4\,kpc. At this luminosity the source is expected to be accreting at the super-critical regime, with the expected value of $L_{\rm crit}$ at $\sim$ $2-4\times10^{36}$ erg s$^{-1}$ \citep[see][assuming the same distance]{2016MNRAS.458.2745H}. At this regime, the source is expected to transform from a plasma-dominated regime with a dominant pencil-like emission beaming pattern to a radiation dominated regime with a dominant fan-like emission beaming pattern \citep{basko-sunyaev1976,becker2012,2015MNRAS.447.1847M}. The complex pulse profile morphology and its energy dependence in \src\, is however difficult to be identified with a pure pencil-like or fan-like beaming pattern. The CRSF parameters also exhibit strong pulse phase dependence but effects like  lightbending, relativistic beaming, reflection, and the mixing of two columns  makes the interpretation of the CRSF parameters very difficult. While the results indicate a overall complex beaming geometry, a detailed modeling of the energy dependence of the pulse profiles together with the pulse phase dependence of the CRSF parameters is required to map the complex variations, a.k.a the accretion geometry in this source. We leave this for a future work.

\section*{Acknowledgments}
The authors would like to thank the referee for his/her constructive comments and suggestions that improved the contents of the paper.
The research work at Raman Research Institute is funded by the Department of Science and Technology, Government of India. This publication use data from the ASTROSAT mission of the Indian
Space Research Organisation (ISRO), archived at the Indian Space Science Data Center (ISSDC). We would like to thank Sreenandini for her help with initial data handling.

\bibliography{1538_bibtex}{}
\bibliographystyle{mnras}
\end{document}